| | |
|---|---|
| **Title** | **Gas sensing properties of multiwall carbon nanotubes decorated with rhodium nanoparticles** |
| Authors | R. Leghrib[a], T. Dufour[b], F. Demoisson[b,c], N. Claessens[b], F. Reniers[b], E. Llobet[a] |
| Affiliations | [a] MINOS - EMaS, University Rovira i Virgili, 26, Països Catalans, 43007 Tarragona, Spain <br> [b] Service de Chimie Analytique et Chimie des Interfaces (CHANI), Université Libre de Bruxelles, Faculté des Sciences, CP255, Boulevard du Triomphe 2, B-1050 Bruxelles, Belgium <br> [c] Laboratoire Interdisciplinaire Carnot de Bourgogne (ICB), Département Nanosciences - Equipe MaNaPI: UMR 5209 CNRS/Université de Bourgogne, 9 Avenue Alain Savary, BP 47 870, F-21078 Dijon Cedex, France |
| Ref. | Sensors and actuators B: Chemical, 2011, Vol. 160, Issue 1, 974-980 |
| DOI | http://dx.doi.org/10.1016/j.snb.2011.09.014 |
| Abstract | In the present work, multiwalled carbon nanotubes were decorated with rhodium nanoparticles using a colloidal solution in the post-discharge of an RF atmospheric plasma of argon (Ar) or argon/oxygen (Ar:O$_2$). The properties of these hybrid materials towards the room temperature detection of NO$_2$, C$_2$H$_4$, CO, C$_6$H$_6$ and moisture were investigated and discussed in view of compositional and morphological studies. It was found that the presence of oxygen in the plasma treatment is essential to significantly enhance the gas response of Rh-decorated multiwalled carbon nanotubes and to avoid response saturation even at low gas/vapor concentrations. These desirable effects are attributed to the presence of oxygen during the CNT plasma treatment since oxygenated vacancies act both as active adsorption sites for gases and as anchoring sites for Rh nanoparticles (the presence of Rh nanoclusters is nearly doubled in Ar:O$_2$ treated samples as compared to Ar treated samples). The oxygen treatment also makes easier the charge transfer between Rh nanoparticles and carbon nanotubes upon gas adsorption. The method for treating and decorating multiwalled carbon nanotubes used here is simple, fast and scalable for producing gas sensitive nanohybrid materials with uniform and well-controlled properties. |

# 1. Introduction

Carbon nanotubes (CNTs) have been widely studied in different applications of nanotechnology thanks to their interesting physico-chemical properties [1–3]. Taking into account their chemical stability and large surface area to volume ratio, CNTs have been considered as new supports for metal catalysts [4,5]. Since the adsorption of a small quantity of chemical species can result in a dramatic change of their conductivity, CNTs are suited to detect gases at low concentrations (e.g. low ppb level) [6]. Furthermore, the possibility to operate CNT gas sensors at room temperature makes them very interesting to integrate low power consumption detectors [7]. A gas sensor can be built using a simple transducer (comb electrodes) to monitor the electrical resistance of a CNT-based film.

In recent years, numerous theoretical and experimental studies have been focused on the functionalization of CNT surface and/or the decoration of their sidewalls with metal nanoclusters, in a quest for enhancing sensitivity and ameliorating selectivity to different toxic gases or vapors [7–13]. The modification of CNT surfaces with metal nanoparticles is of interest for gas sensing since metal nanoclusters have a broad range of diverse structures and can provide a wide range of reactivity with different species. Indeed, the concept of using CNT–metal cluster hybrids as the sensitive material of a device where the metal cluster surfaces act as reactive sites for the adsorption of the target molecules was theoretically introduced in [11]. The first researchers that experimentally exploited this concept were Kong et al. [8], Kumar and Ramaprabhu [12] and Star et al. [13], who used Pd-decorated CNTs or Pt decorated CNTs to detect hydrogen, and an array of single wall CNTs decorated with different metals (Rh, Ni, Au, Pd, etc.) for detecting H$_2$S, CH$_4$, H$_2$, CO and NH$_3$, respectively. Recently [7], we have shown the selective detection of low C$_6$H$_6$ concentrations employing an array of Rh, Ni, Au, and Pd decorated multiwall carbon nanotube (MWCNT) gas sensors operated at room temperature.

Among the different metals one can consider for decorating carbon nanotube sidewalls, rhodium is of interest since the Rh-MWCNT catalyst has been reported to be effective in the hydrogenation of aromatic rings at room temperature. In addition, the catalyst can be reused at least six times without losing catalytic activity for arene hydrogenation and its catalytic activity is much higher than that from commercially available Rh nanocatalysts [14,15]. Although there is a weak interaction between molecular benzene and CNTs [16], the presence of Rh nanoparticles on the surface of MWCNTs helps enhancing the sensor signal upon benzene adsorption (i.e., among the different metals studied, Rh led to the highest benzene response) [7].

In this paper, the effects of using two different atmospheric plasmas (i.e., Ar or an Ar–$O_2$ mixture) while preparing Rh decorated multiwall carbon nanotubes on the sensing properties of the resulting nanohybrid materials are studied. The performance of gas sensors employing such materials are compared in terms of their response towards $NO_2$, CO, $C_2H_4$, $C_6H_6$ and moisture. The differences in their response are discussed in view of the data gathered by Transmission Electronic Microscopy (TEM) and X-ray Photoelectron Spectroscopy (XPS) analyses. Combined, the gas sensing, compositional and morphological studies give insight about the role of oxygen species grafted to CNTs during the plasma treatment both for ameliorating the decoration with metal nanoparticles and enhancing the gas response of the resulting nanohybrid material.

## 2. Experimental

### 2.1. Hybrid nanomaterials: synthesis and characterization

The multiwall carbon nanotubes used were provided by Nanocyl S.A. They were grown by Chemical Vapor Deposition with purity higher than 95%. Nanotubes were up to 50 m in length and their outer and inner diameters ranged from 3 to 15 nm and 3 to 7 nm, respectively. These nanotubes were decorated with Rh nanoparticles employing a colloid solution supplied by the Department of Chemistry from the University of California at Berkeley (USA). For this solution the capping agent was poly(vinylpyrrolidone) (PVP) and the solvent was ethanol. Solution concentration was 4 mmol/L, which represented 0.31 g $L^{-1}$ of Rh. The size of Rh nanoparticles was 8.8±1.7 nm [17]. The plasma activation and deposition processes are further explained in [18,19]. At first, the experimental process required the cleaning of carbon nanotubes for 15 min in a methanol solution. Surface treatment and metal decoration were conducted using plasma post-discharge at atmospheric pressure. The plasma was generated with a RF torch (Atomflo-250, from Surfx Technologies LLC), powered at 80 W and working for an argon flow of 30 L/min. For some of the samples prepared, oxygen was mixed to the argon gas at the rate of 20 mL/min. Carbon nanotubes, placed 10 mm away from the RF torch were exposed to the plasma post-discharge for 2 min to activate their surface. Then, during 30 s, the rhodium colloid solution was sprayed in the plasma post-discharge. Treated nanotubes had to be exposed 3 more minutes before switching off the RF torch. As a final step, they were introduced for 5 min into a methanol solution under ultrasonication.

The chemical composition and the surface morphology of the different hybrid materials employed as active layers of the sensors fabricated were analyzed by X-ray photoelectron spectroscopy (XPS) and transmission electron microscopy (TEM), respectively. The results of these chemical and morphological studies can be found in [18].

## 2.2. Sensor fabrication and characterization

Sensor substrates consisted of an integrated array of four inert micro-hotplate substrates, each one comprising a polysilicon heating resistor sandwiched between two thin silicon nitride layers and a pair of interdigitated Pt electrodes on top (the electrode gap was 50 µm). The membrane thickness was about 0.6 µm and the active (hot) area of 500 µm×500 µm [20]. Plasma-treated Rh-decorated MWCNT were dispersed in dimethyl formamide (DMF) by ultrasonication at room temperature to form uniform suspensions. The suspensions were airbrushed onto the interdigitated electrodes. After DMF evaporated (sensors are kept on a hotplate during the airbrushing process), a mat of metal decorated CNTs spanned across neighboring fingers. By carefully controlling its process parameters (e.g., nozzle to substrate distance, substrate temperature, carrier flow and amount of CNTs within a suspension), highly reproducible mats of Rh-decorated CNTs are obtained. Finally, sensors were annealed at 250°C for 4 h in an air flow to completely remove the remnant DMF and improve the nanotube-electrode contact.

The gas sensing properties at room temperature of Rh-decorated carbon nanotubes treated in Ar or Ar+$O_2$ plasmas were tested in the presence of $NO_2$, $C_2H_4$, CO, and $C_6H_6$. Humidity response was also investigated. To perform the measurements, the gas sensors were placed in a 2×10−5m3 volume Teflon/stainless steel test chamber. The desired concentrations of each species were obtained using a computer controlled measurement rig that employed mass flow meters and calibrated gas bottles. Dry air was used as carrier gas. R.H. in the measurement rig was 10% at 30°C. The concentrations tested were as follows: 50, 100, 200, 500 and 1000 ppb for benzene vapors; 50, 100, 500 and 1000 ppb for $NO_2$; 3, 7, 15 and 30 ppm for $C_2H_4$; and 2, 5, 10 and 20 ppm for CO. Moisture responses to changes in R.H. from 10 to 50%, 50% to 80% and 10% to 80% were measured. Finally, to better study moisture effects on gas response, 50 and 100 ppb of $NO_2$; 2 and 5 ppm of CO; and 50 and 100 ppb of benzene were measured at 50% R.H. To assess the reproducibility of results, each measurement was replicated 4 times. An Agilent 34970A multimeter was used for continuously monitoring the electrical resistance of the sensors. DC measurements were performed throughout this study. The measurement process was as follows (identical for all species tested): the air flow was set to 100 sccm and kept constant. Data acquisition started 20 min before injecting a continuous flow of the desired concentration of a given species into the measurement chamber. 20 min after the injection, the sensors were flushed with dry air and their temperature was raised to 150°C for promoting the cleaning of their surface. 20 min after heating had been switched off, a new measurement was performed.

## 3. Results and discussion

### 3.1. Effect of plasma treatment conditions on the nanohybrid materials

XPS and TEM results show that the introduction of oxygen in the plasma used to treat the surface of CNTs induces the grafting of a higher amount of Rh nanoparticles onto CNT sidewalls [18]. XPS confirms that adding 20 mL/min of oxygen to the Ar plasma increases the amount of Rh from 1.2% (in wt.%) in Ar treated samples to 2.2% (in wt.%) in Ar-$O_2$ treated samples. Moreover, the oxidation state of the grafted Rh remains unchanged. TEM analysis shows that this method preserves the original size of the Rh nanoparticles from the colloidal solution. These results can be found elsewhere [18]. Additionally Suarez-Martinez et al. showed in [21] that the inclusion of oxygen in an RF plasma treatment grafts oxygen species to the CNT surface, which helps creating nucleation/anchoring sites for metal clusters. As a result, more Rh nanoparticles are attached to CNT sidewalls and with more uniform distribution than on pristine tubes or on tubes treated with Ar plasma only. In [18], it is suggested that the introduction of oxygen during CNT plasma treatment favors the decoration with Rh nanoparticles because these are immobilized at oxygenated vacancies present in CNT sidewalls and the capping agent of colloidal particles is activated, which helps forming a bridge between C and Rh atoms. Giordano et al. [22] found that the oxidation of the outer surface of multi-walled carbon nanotubes by nitric acid leads to the formation of carboxylic groups that can be used to anchor rhodium carbonyl moieties. Thus, the pretreatment of CNTs in oxygen plasma is very necessary to get a high amount of metal particles attached onto nanotube surface with good homogeneity [18], which are of paramount importance for getting high reactivity to gases of the resulting hybrid material. The amounts of Rh grafted on CNT sidewalls are highly reproducible provided that the plasma parameters are kept the same. Furthermore, the functionalization of CNT surface with oxygen species prior to metal decoration affects the density of states of valence bands [23] and this improves the exchange of electrons at the interface between CNTs and metal nanoparticles, which in this way increases the sensitivity of the material.

## 3.2. Gas sensing analysis

After exposing the sensors based on Rh-CNTs treated in argon or argon/oxygen plasmas to the different gases studied, it was found that sensors showed different reactivities depending on the material treatment and the gas detected. Fig. 1 shows the typical responses, at room temperature, of our hybrid materials towards $NO_2$, $C_6H_6$ and CO. Although sensors can recover their baseline by cleaning them with air at room temperature, heating at 150°C speeds up sensor baseline recovery. The slow recovery of baseline resistance at room temperature and the effect of heating at 150°C are shown in Fig. 2. This moderate temperature helps desorbing the species adsorbed on the surface of the hybrid materials. This is supported by the fact that their response to the different species tested decreases if the sensors are operated at 150°C. This effect is confirmed by DFT calculations performed on defective CNTs [24]. Saturation in the response is seen for CO concentrations higher than 5 ppm (see Fig. 1e in which if CO concentration is further increased to 10 or 20 ppm, sensor resistance variations remain below the noise level). If response time is defined as the time needed to reach 90% of the steady state resistance value after gas concentration is changed stepwise, in Rh-decorated CNT sensors response time is below 10 min for CO, $C_6H_6$ and $C_2H_4$ and near 20 min for $NO_2$.

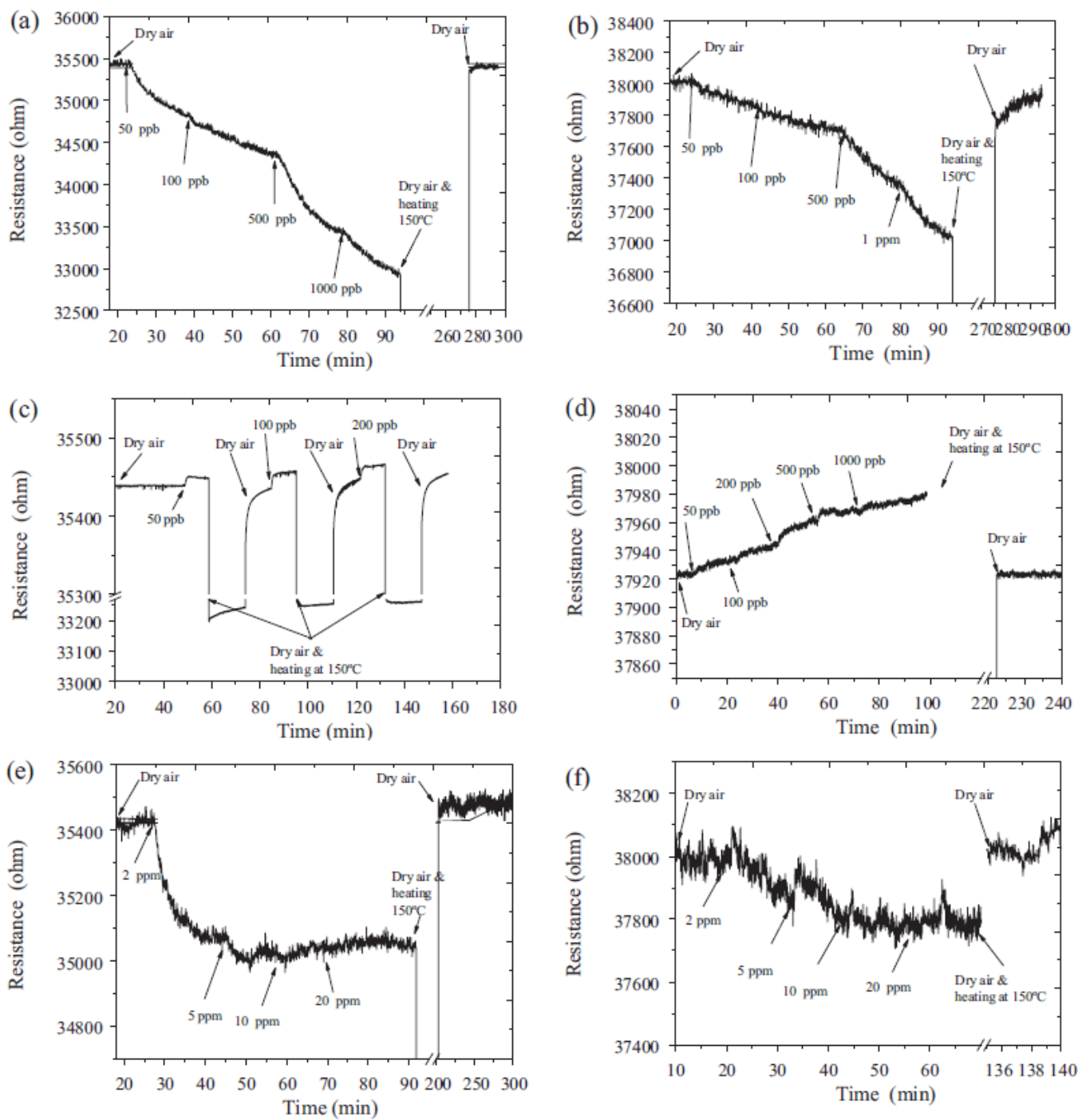

Fig. 1. Typical room-temperature responses of sensors employing plasma-treated Rh-CNTs to nitrogen dioxide: (a) Ar:$O_2$ plasma, (b) Ar plasma; benzene: (c) Ar:$O_2$ plasma, (d) Ar plasma and carbon monoxide: (e) Ar:$O_2$ plasma, (f) Ar plasma.

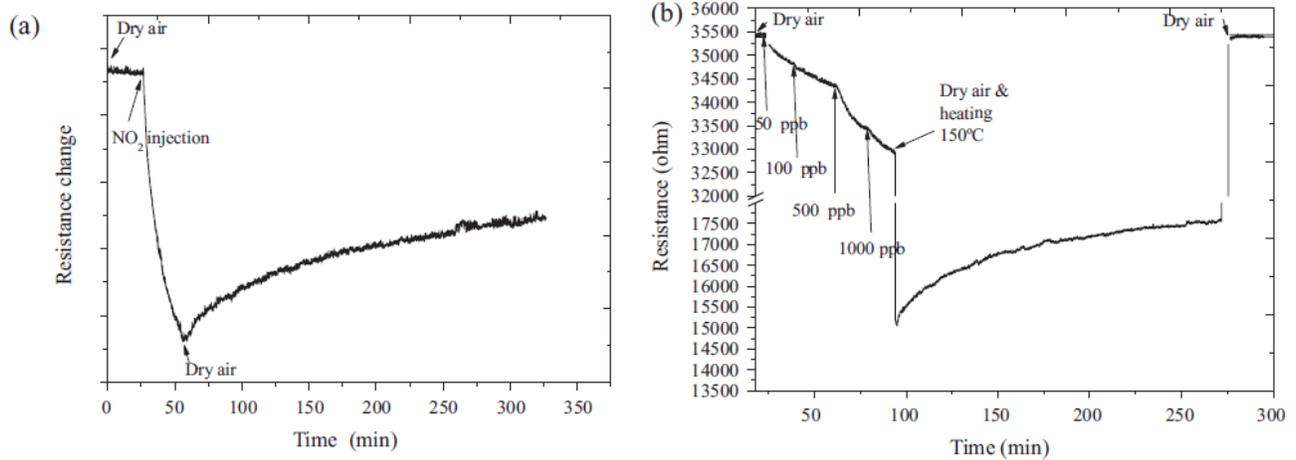

Fig. 2. (a) Response and slow recovery of a Rh-decorated CNT sensor at room temperature and (b) room temperature response and complete baseline recovery of a Rh-decorated CNT sensor when heating is applied during the cleaning cycle.

The subplots in Fig. 3 show the response of the two hybrid materials (i.e., Rh-MWCNT treated in Ar or in Ar:$O_2$ plasmas) towards different concentrations of $C_2H_4$, CO, $NO_2$ and $C_6H_6$. Response is defined as $(Rg-Ra)/Ra$, where Ra is the sensor resistance in clean air (i.e. baseline resistance) and Rg is the resistance in the presence of a given species. Considering the species measured (see Fig. 3a–d), for both materials and for the different concentrations tested, the absolute value of the response is higher for $NO_2$ followed by $C_2H_4$, CO and finally $C_6H_6$. Rh-MWCNTs treated in argon plasma show a clear saturation effect in their response that appears, in some cases, even at the lowest concentrations measured. On the other hand, Rh-MWCNTs treated in argon–oxygen plasma show good linearity in their response to $NO_2$ and $C_6H_6$. However, for $C_2H_4$ and CO their response saturates at 2 and 10 ppm, respectively.

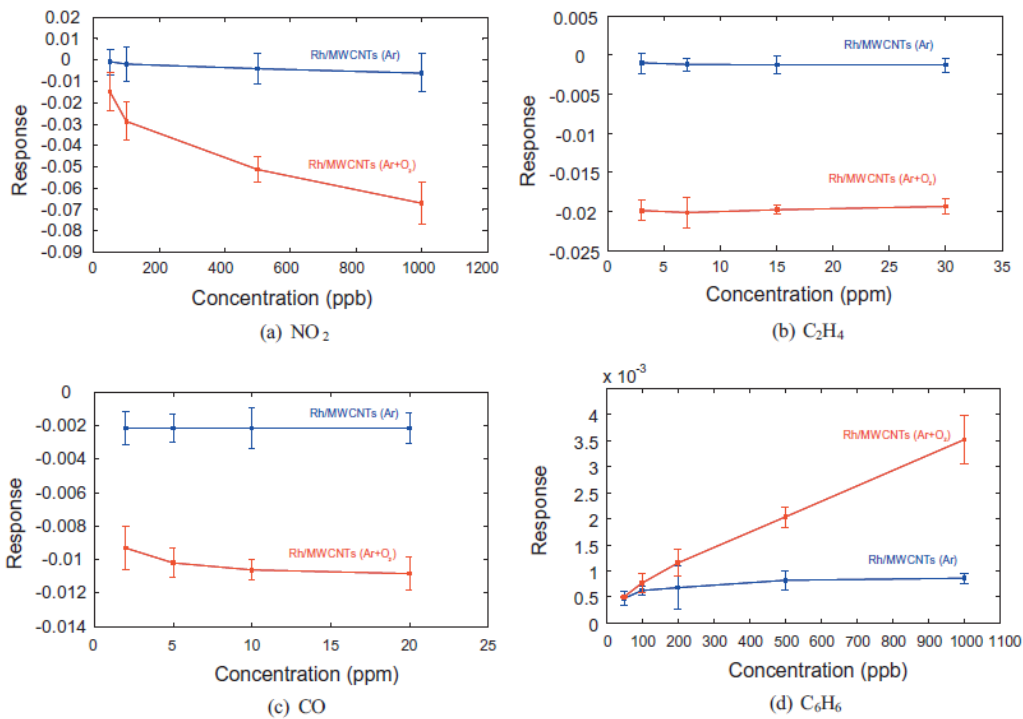

Fig. 3. Room-temperature response and associated error bars of sensors employing Rh-CNTs either treated in argon or argon–oxygen plasmas to (a) nitrogen dioxide, (b) ethylene, (c) carbon monoxide and (d) benzene.

Fig. 3 shows that hybrid materials treated in Ar plasma show lower response to the gases tested than materials treated in Ar:$O_2$ plasma. The increase in response achieved by the treatment under an Ar:$O_2$ plasma is due to both the increased amount of Rh nanoparticles grafted to the outer surface of CNTs and to the creation of oxygenated vacancies, which can act as active adsorption sites for gas molecules [25]. The early saturation in response experienced with materials treated in argon plasma only could be due to their lower coverage with Rh nanoparticles and lower amount of oxygenated defects. The fact that the saturation in response is also observed in Ar:$O_2$ plasma treated samples in the particular case of ethylene could be due to the non-suitability of Rh for detecting this species [7].

These results determine the dominant interaction mechanism between the gases tested and the metal decorated CNTs. This mechanism is the charge transfer between adsorbed gas molecules and CNTs via Rh nanoparticles. This alters the conductivity of the hybrid material. The direction of this charge transfer depends on the oxidizing or reducing nature of the gas environment [26]. Recently, Zanolli and co-workers published a deep combined experimental and theoretical study on the sensing properties of Au decorated CNTs towards benzene, carbon monoxide and nitrogen dioxide [27].

In this study, nitrogen dioxide is predicted to accept electronic charge from the Au-CNT system, which is consistent with the electron acceptor character of that molecule. The nature of the charge transfer between $NO_2$ and Au-CNT indicates that holes are injected in the metal decorated carbon nanotube strengthening its p-type semiconductor behavior and lowering further the Fermi level of the system. This explains the decrease in resistance measured in Au-CNT sensors and also applies to the Rh-MWCNT sensors studied here.

On the other hand, the interactions of Au-CNTs with benzene are so weak that no significant charge transfer could be computed, suggesting that Au-CNTs are not suitable for detecting benzene [27]. A metal such as Rh, which shows a stronger interaction than Au with CNTs is more suitable for detecting benzene since the detection mechanism relies on the biding of the molecule to the nanoparticle. Benzene is well-known to be an electron donor. Upon benzene adsorption on Rh nanoparticles, electrons are injected into the Rh-CNT system decreasing the concentration of free carriers (i.e., holes), which explains the increase in resistance measured in Rh-MWCNT samples.

The adsorption of a CO molecule on Au-CNT results in a small fraction of electron charge transferred towards CO [27]. The interaction between CO and Au nanoclusters is quite complex [28] since it strongly depends on the overlap of the 5σ and 2π* orbitals of the CO with the d orbitals of the gold nanoparticle, resulting in a mechanism of electron donation/back-donation (i.e., CO donates electrons from its σ orbitals to the $d_σ$ orbitals of gold and the filled $d_π$ orbitals of Au donate electrons back into the empty π orbitals of CO). A similar mechanism could explain the decrease in resistance observed for Rh-MWCNT sensors in the presence of CO.

The operational principle in many ceramic humidity sensors is physisorption of water molecules on an initially chemisorbed hydroxyl layer. Water physisorption leads to protonic conduction on the surface, which increases the overall conductance of the sensor [29]. The process of physisorption onto a hydroxyl layer cannot explain the present results since Rh-MWCNT sensors decrease their conductance when humidity increases. These experimental results are in agreement with those previously reported by Varghese et al. in [30], in which DC and AC measurements ruled out water physisorption as the operational principle in MWCNTs. Instead, they suggested that water vapor directly interacts with p-type MWCNTs donating electrons into the valence band and thus, decreasing the number of holes. Additionally, in a recently published paper [24], Zanolli and co-workers have reported computational studies on the interaction of water vapor with defective carbon nanotubes (e.g., CNTs with oxygenated vacancies). Water vapor is predicted to mildly interact with defective tubes in which a small charge transfer occurs between adsorbed water molecules and CNTs.

Similar studies should be performed to elucidate the interaction of ethylene molecules with metal decorated CNTs. Fig. 4 shows the response to moisture of the different sensing hybrid materials. Three different cases were considered: the responses when relative humidity changed from 10 to 50%, from 10 to 80%, and from 50 to 80% were measured. Hybrid materials treated in Ar:$O_2$ plasma show a significantly higher humidity response than materials treated in Ar plasma only. In fact, another effect of introducing oxygen during the plasma treatment of CNTs is the modification of their surface polarity resulting in increased hydrophilicity, which enhances the response to moisture. Fig. 4 shows that there is a moderate response when relative humidity changes from 50 to 80%. Furthermore, a sudden 30% change in relative humidity is quite unlikely. Therefore, this fact suggests that the material would suffer from moderate humidity interference provided that sensors are operated in an ambient atmosphere where moisture remains within this range. Fig. 5 compares the response of an Ar:$O_2$ plasma treated Rh-MWCNT sensor to nitrogen dioxide, carbon monoxide and benzene when relative humidity was set either to 10% or 50%. This study was performed for oxygen treated sensors because these were the most responsive to the different species tested. Moisture levels were set to 10% and 50% because sensors significantly responded to humidity changes in this range (see Fig. 4). The absolute values of the slopes of the curves shown in Fig. 5 decrease when humidity increases. In other words, the sensitivity towards these species slightly diminishes when background moisture is raised from 10% to 50%. These results seem to indicate that moisture and the different gases and vapors tested compete for adsorption sites at the surface of CNTs and Rh clusters. The use of a humidity sensor should be considered for compensating the effects of small humidity changes.

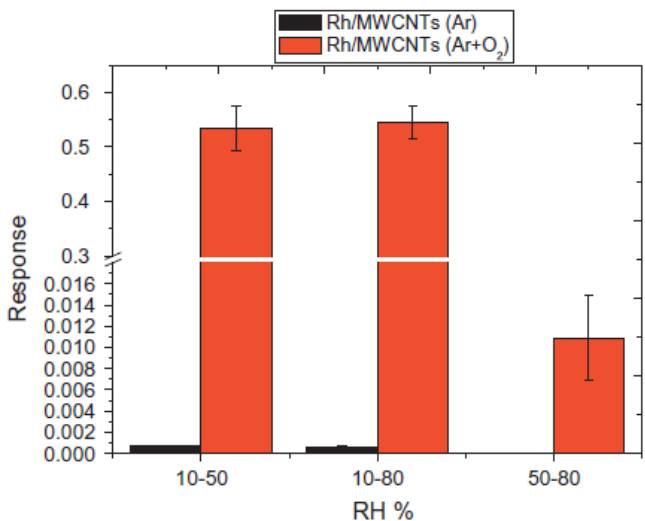

*Fig. 4. Room-temperature response to moisture changes of sensors employing Rh-CNTs either treated in argon or argon–oxygen plasmas.*

*Fig. 5. Room-temperature response to nitrogen dioxide, carbon monoxide and benzene of sensors employing Ar:$O_2$ treated, Rh CNTs when relative humidity (measured at 30 °C) is either 10% or 50%.*

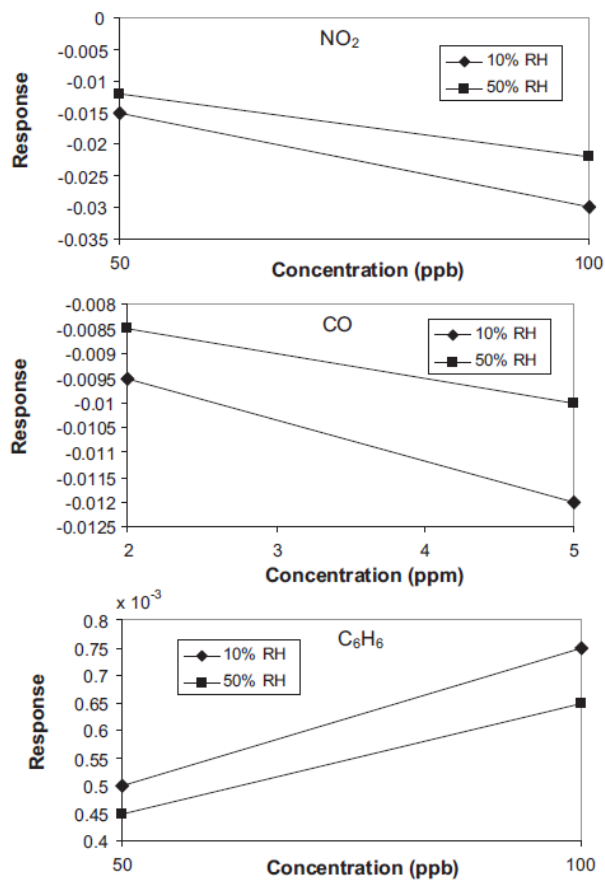

According to the number of species studied, different concentrations tested and number of replicate measurements performed, every sensor characterized underwent 80 detection-recovery cycles, which spanned for about one month. No deactivation of Rh nanoparticle activity (i.e., significant response drift or response losses) was observed.

# 4. Conclusions

The introduction of oxygen during the plasma treatment of carbon nanotubes leads to the creation of active sites (i.e. oxygenated vacancies) on the surface of CNTs. Additionally, reactive oxygen activates the capping agent of colloidal nanoparticles. Combined, these two effects improve the metal nanoparticle–CNT surface interaction, which results in a higher amount of metal nanoparticles grafted to CNT sidewalls and a more homogeneous decoration of CNTs. For example, the inclusion of oxygen in the Ar plasma leads to nearly doubling the weight of Rh attached to CNTs and preserves the original size of the nanoparticles from the colloidal solution. Furthermore, oxidative treatments affect the density of states of valence bands and increase the work function of MWCNTs making it closer to that of metals such as Rh. As a result, it is easier for electrons to travel between the metal nanoparticles and the CNT, with the direction of charge transfer depending on the gaseous environment.

Therefore, the response of oxygen plasma treated, Rh-decorated MWCNTs to different gases and vapors (e.g. $NO_2$, $C_2H_4$, CO or $C_6H_6$) is substantially increased. It is worth stressing that the saturation in the response observed with Ar plasma treated, Rh-decorated MWCNTs is solved when oxygen is included in the plasma treatment. The inclusion of oxygen during the plasma treatment turns CNTs hydrophilic and this has an impact in their response to ambient moisture. However, the response of the oxygen–argon plasma treated hybrid materials to humidity changes in the 50 to 80% R.H. range remains moderate.

The CNT treatment and metal decoration employed here uses an atmospheric plasma source. It is a simple, fast and scalable method for producing gas sensitive nanohybrid materials with uniform and well-controlled properties.

# 5. Acknowledgments

This work was supported in part by the European Commission under the FP6-NMP grant Nano2hybrids (no. 033311 STREP), and the URV PhD fellowship program. We are thankful to Dr. Isabel Gràcia and Dr. C. Cané from the CNM (IMB-CSIC, Spain) for providing us with the microhotplate transducers and to Prof. Gabor A. Somorjai from the Department of Chemistry (University of California) in Berkeley, for supplying the colloidal solutions of rhodium.